\documentclass[aps,prb,twocolumn,groupedaddress,floatfix,showpacs]{revtex4-2}

\usepackage[T1]{fontenc}
\usepackage{graphicx}
\usepackage{amsmath}
\usepackage{color}
\begin{document}

\title{Numerical analysis of grasshopper escapement}

\author{David Ziemkiewicz}
\email{david.ziemkiewicz@utp.edu.pl}

 \affiliation{Institute of
Mathematics and Physics, UTP University of Science and Technology,
\\Aleje Prof. S. Kaliskiego 7, 85-789 Bydgoszcz, Poland.}


\begin{abstract}
The dynamics of driven, damped pendulum as used in mechanical clocks is numerically investigated. In addition to the analysis of a well-known mechanisms such as chronometer escapement, the unusual properties of Harrison's grasshopper escapement are explored, giving some insights regarding the dynamics of this system. Both steady state operation and transient effects are discussed, indicating the optimal condition for stable long-term clock accuracy. Possibility of chaotic motion is investigated.  
\end{abstract}
\maketitle

\section{Introduction}
A weight-driven pendulum clock is a nonlinear, dynamic system consisting of a damped, driven pendulum and the so-called escapement mechanism. The role of the mechanism is twofold. It regulates the speed of the clock by binding it to the period of the pendulum. The second function is to provide the energy to the pendulum, so that a nonzero oscillation amplitude can be maintained despite the friction. The energy is added by pushing the pendulum along some part of its motion, geometrically determined by escapement construction. Thus, the frequency of the force is always the same as the pendulum itself, making it a self-excited oscillator \cite{Denny}. Extensive overview of the physics of pendulum clocks is presented in \cite{Rawlings}.

The long, historical struggle to attain highest possible accuracy (stability of the period) consists of devising new ways to separate the pendulum from various disturbances, including the effect of escapement itself; the same mechanism that keeps the pendulum in motion is also responsible for instability. Accuracy of many popular escapement types has been extensively analysed; some recent studies include dead-beat escapement \cite{Kesteven}, gravity escapement \cite{Kesteven} and anchor escapement \cite{Denny,Headrick}. Some research covers mechanisms of historical significance, but relatively poor timekeeping properties such as verge and foliot \cite{Blumenthal,Danese}. The closely related field of dynamics of mechanical watches is still dynamically developing \cite{Xu}.
By moving from an idealized pendulum to physical one, a large number of factors need to be taken into account \cite{Nelson}, which can be roughly divided into three categories depending on the source of error: pendulum, environment and escapement mechanism. The focus of this paper is the interaction between pendulum circular error present for any nonzero swing angle and escapement error caused by the disturbance of the pendulum motion by the clock mechanism.
  
In \cite{Hoyng} the author used an approach that averages the force over period to study the long-term effects on the pendulum, on a time scales much longer than single period. The method of averaging is a well-developed approach to analyse periodic, nonlinear systems \cite{Libre}. On the other hand, recent advancements in computer technology make a direct numerical integration of the partial differential equations of pendulum motion with resolution much smaller than single period a practical approach, even for extended timescales spanning weeks or months \cite{Hoyng}. The advantage of such calculation lies in its flexibility and amount of provided data; the fine structure of the motion on a time scale smaller than the period is preserved. Additionally, transient effects can be readily studied.

Finally, chaotic dynamics of the whole pendulum-escapement system is explored. The pendulum is an excellent tool to study chaotic motion; such a behaviour may emerge in systems with mass suspended on elastic string \cite{Arinstein}, periodically driven \cite{Kobes}, parametrically damped \cite{Smith} and double \cite{Levien} pendulums. Strongly damped pendulums may exhibit symmetry breaking \cite{Isohatala}. In a clock, the mechanical oscillations of the mechanism coupled to the pendulum provide a nonlinear system with chaotic dynamics \cite{Moon}.

\section{Circular error}
\begin{figure}[ht!]
\centering
\includegraphics[width=.6\linewidth]{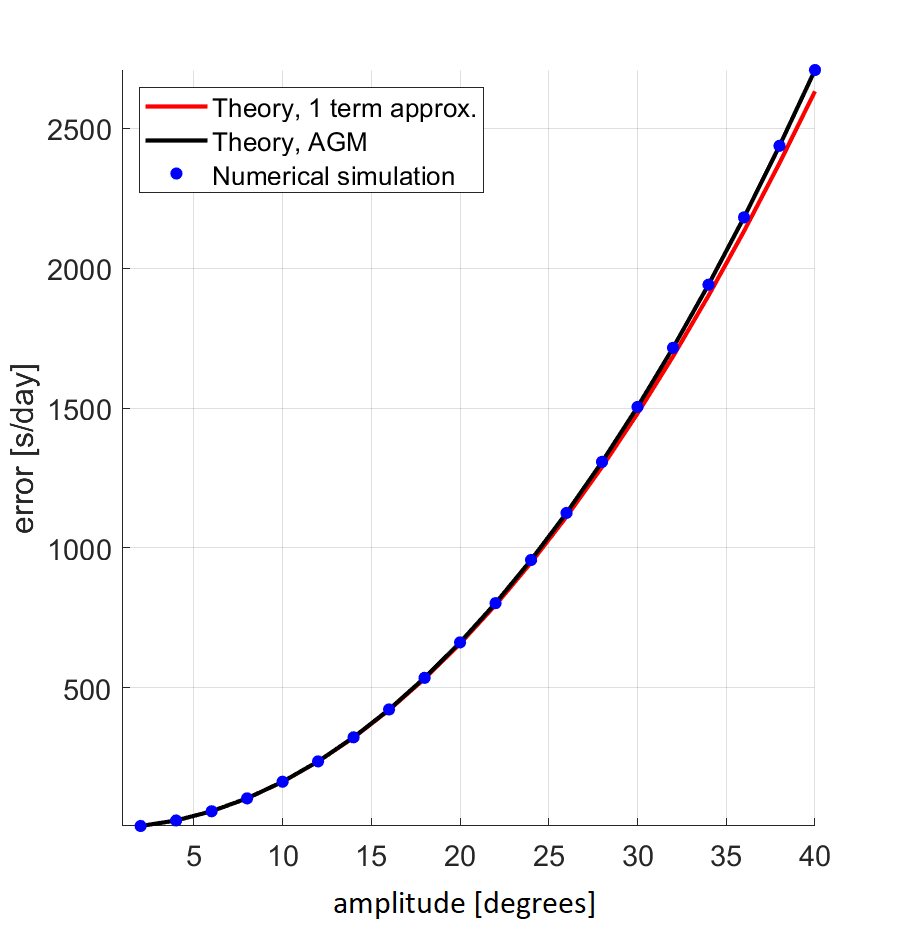}
\caption{The analytically calculated and numerically simulated pendulum period as a function of amplitude}\label{fig:1}
\end{figure}

The equation of motion for a pendulum modelled as a point mass $m$, suspended on a weightless string with length $L$ is
\begin{equation}\label{eq:motion1}
m\ddot{\alpha}-\gamma\dot{\alpha}=\frac{M}{I}=\frac{g}{L}\sin(\alpha).
\end{equation}
where $\alpha$ is the pendulum angle, $\gamma$ is a damping constant, $M=mgL\sin(\alpha)$ is the total moment of force and $I=mL^2$ is the moment of inertia. In a physical pendulum, $M$ and $I$ have a more complicated forms. By substituting $\sin(\alpha)\approx\alpha$, one obtains an equation for harmonic oscillator with period $T_0=2\pi\sqrt{L/g}$ and angle $\alpha=A\sin(\omega t + \phi_0)$, where $\omega=2\pi/T$, $A$ is the amplitude and $\phi_0$ is the starting phase. Exact solution of Eq. (\ref{eq:motion1}) involves a complete elliptic integral \cite{Carwalhaes}. The period of the pendulum $T$ is always larger than $T_0$ and the difference is called the circular error. The value of $T$ can be expressed as a series \cite{Bishop}
\begin{equation}\label{eq:circ}
T=T_0\left[1+\left(\frac{1}{2}\right)^2\sin^2\frac{A}{2}+\left(\frac{1\cdot3}{2\cdot4}\right)^2\sin^4\frac{A}{2}+\dots\right].
\end{equation}
As a first approximation, the error is proportional to the square of amplitude. Another approach, which will be used in this manuscript, is the Arithmetic-Geometric mean (AGM) \cite{Carwalhaes}, which provides particularly accurate estimation of circular error. Both methods are shown on the Fig. \ref{fig:1} along the results of numerical simulation of pendulum motion.
There is an excellent agreement between the theory and simulation results. Detailed simulation description and accuracy estimations are presented in the Appendix A. 
Naturally, to minimize the influence of circular error one should aim for the smallest practical swing angle. However, even for $A=2$ degrees, the circular error $E_C \sim 7\cdot10^{-5}$ which corresponds to a non-trivial change of rate on the order of 6 seconds/day. One of the ways of correcting the circular arc is modifying the pendulum suspension in such a way that it follows a cycloid instead of circular arc, as shown by Hyugens in 1673 \cite{Cidoncha}. Interestingly, as it will be shown, the elimination of circular error is not always beneficial to the clock accuracy.

\section{Escapement error}
Any physical pendulum is subject to friction, which causes reduction of amplitude over time. The common parameter describing the energy loss of an oscillator is the Q factor
\begin{equation}\label{eq:Q}
Q = \frac{2\pi E}{\Delta E} = \frac{m\omega}{\gamma},
\end{equation}
where $E$ is the total energy and $\Delta E$ is the energy lost in one period. In order to sustain pendulum motion, the escapement mechanism of a clock needs to provide the lost energy $\Delta E$ in every period by doing work
\begin{equation}\label{eq:work}
\Delta E = \int\limits_{\alpha_1}^{\alpha_2} M_F(\alpha)d\alpha
\end{equation}
where $M_F$ is the moment of force (torque) exerted on the pendulum and $\langle\alpha_1,\alpha_2\rangle$ is the range of pendulum angle where the force is applied. However, any force acting upon the pendulum will change its rate, introducing an escapement error $E_E$. 
\begin{equation}\label{eq:EE}
E_E = \frac{T'-T}{T},
\end{equation}
where $T$ and $T'$ are periods of the free and disturbed pendulum. With the above convention, positive $E_E$ indicates larger $T'$ and slower clock speed. In general, there are two distinct cases of torque adding energy to the pendulum: 
\begin{itemize}
\item force acting before the pendulum reaches its lowest point $\alpha=0$, directed towards the lowest point,
\item force acting after the pendulum reaches its lowest point $\alpha=0$, directed away from the lowest point.
\end{itemize}
In the first case, the torque adds to the restoring force generated by gravity; therefore, the pendulum acts as though the gravity force was greater and speeds up. Torque applied after $\alpha=0$ reduces the "effective gravity" and slows the pendulum down. One can use a symmetric range of $\alpha$ to avoid influencing the pendulum speed; in other words, the period is constant if the force is an even function of $\alpha$ \cite{Denny}. This is the design goal of a chronometer escapement.

Intuitively, to reduce the error $E_E$ one needs to minimise the force $M_F$. This requirement demands low losses and high Q factor. Such result is in agreement with observations in other systems such as atomic clocks, Q factor is a measure of stability (precision) of an oscillator \cite{Bowden}. However, some amount of friction is necessary for the pendulum to stabilise at some amplitude A. Moreover, as it will be shown, a pendulum with stronger damping reaches the equilibrium point faster, making it more resistant to random external disturbances. Therefore, in the case of mechanical clock, the choice of the optimal Q factor is nontrivial.

\section{Chronometer escapement}
\begin{figure}[ht!]
\centering
a)\includegraphics[width=.8\linewidth]{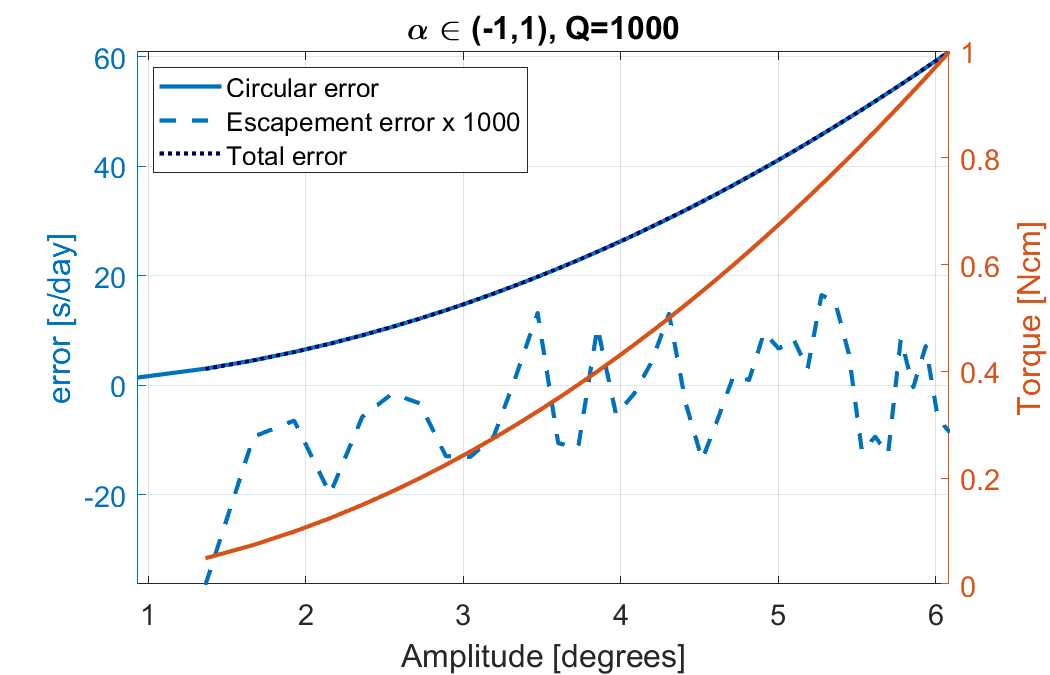}
b)\includegraphics[width=.8\linewidth]{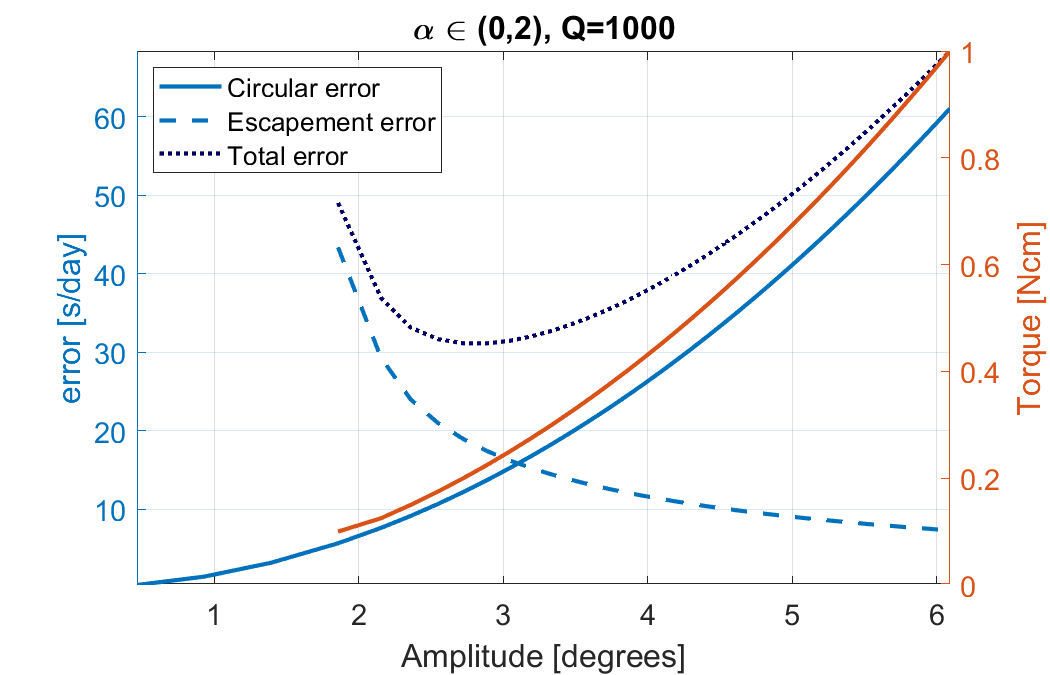}
c)\includegraphics[width=.8\linewidth]{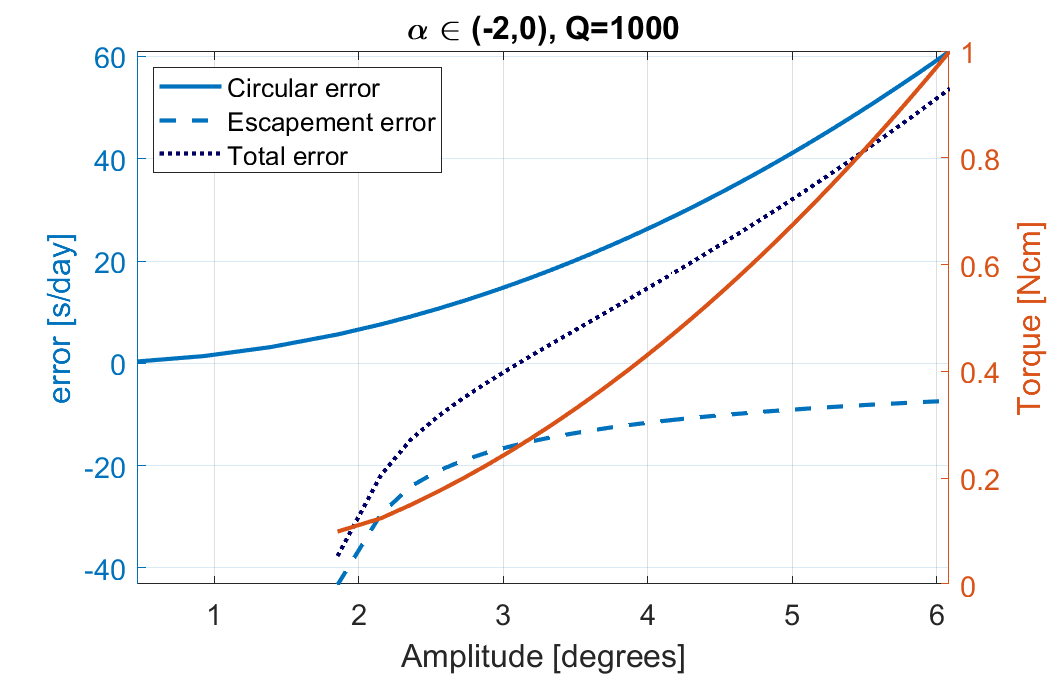}
\caption{Simulation results of chronometer escapement with three different geometries.}\label{fig:2}
\end{figure}

Lets assume a $L=1$ m, $m=1$ kg pendulum driven by a constant torque $M_F(\alpha_1<\alpha<\alpha_2)=\mbox{const}$ applied when the velocity $\dot{\alpha}$ is positive. The friction is assumed to be proportional to speed, e.g. $M_{friction} = L\gamma \dot{\alpha}$, where $\gamma$ is the damping coefficient. Fig. \ref{fig:2} a) shows the typical case where the torque is applied exactly in the middle of the swing. A range of simulations is performed with increasing force $M_F$.
As expected, the escapement error $E_c$ is negligible; the visible, random oscillations on the order of 0.01 seconds/day (relative error $\Delta T/T \sim 10^{-7}$) can be seen as the limit of simulation accuracy. For a given Q factor (here, Q=1000), the energy lost in a single cycle is proportional to the total energy, which is $E \sim m\omega^2A^2$; Therefore, the relation between amplitude and torque is $A^2 \sim M_F$. The total error $E_T=E_C+E_E$ is indistinguishable from circular error. The Fig. \ref{fig:2} b) shows an asymmetric system where the pendulum is pushed when $\alpha>0$. In such a case, the $E_E$ is positive and not negligible. Interestingly, while $E_C$ increases with amplitude, $E_E$ decreases and their sum forms a local minimum. At this point, the change of rate with amplitude, e.g. $\partial E_T/\partial A$ vanishes. This means that the system is locally insensitive to changes of force/amplitude. The particular location of the minimum depends on the ratio of $E_C$ (which is constant) and $E_E$, which depends on the driving force; for a given amplitude $A$, it is a function of the Q factor. One can also notice the lack of data points for $A<\alpha_2$, indicating that the force was insufficient to sustain a stable oscillation. Finally, Fig. \ref{fig:2} c) shows a case where the pendulum is pushed before reaching the equilibrium point. As expected, the escapement error is negative (period is reduced) and the total error has no local minimum. However, $E_T(A)$ is almost linear relation in a large range of $A$, which could be taken advantage of when designing additional error correction mechanism. In both asymmetric cases, the absolute value of $E_E$ decreases with amplitude. This can be explained by the fact that given fixed values of $\alpha_1,\alpha_2$, as the $A$ increases, smaller portion of the swing arc is affected by the escapement. This result is consistent with findings in \cite{Hoyng}, where for a short push taking place at angle $\alpha_1$, the escapement error is $E_E \sim \frac{\tan \alpha_1}{A^2}$. 

\section{Grasshopper escapement}
The grasshopper escapement, invented by John Harrison around 1722, is an interesting mechanism with non-trivial timekeeping characteristics and a large advantage of near-zero sliding friction \cite{Aydlett}. At the first glance, some of its features should be very detrimental to the accuracy; the escapement is pushing the pendulum along its entire path of motion so that it is never free. Moreover, the amplitude is unusually large. Due to these factors, the grasshopper escapement has been historically neglected \cite{Aydlett}. With recent Guinness world record of accuracy within one second in 100 days \cite{McEvoy,Burgess}, there is a renewal of interest in this type of mechanism.

\begin{figure}[ht!]
\includegraphics[width=.8\linewidth]{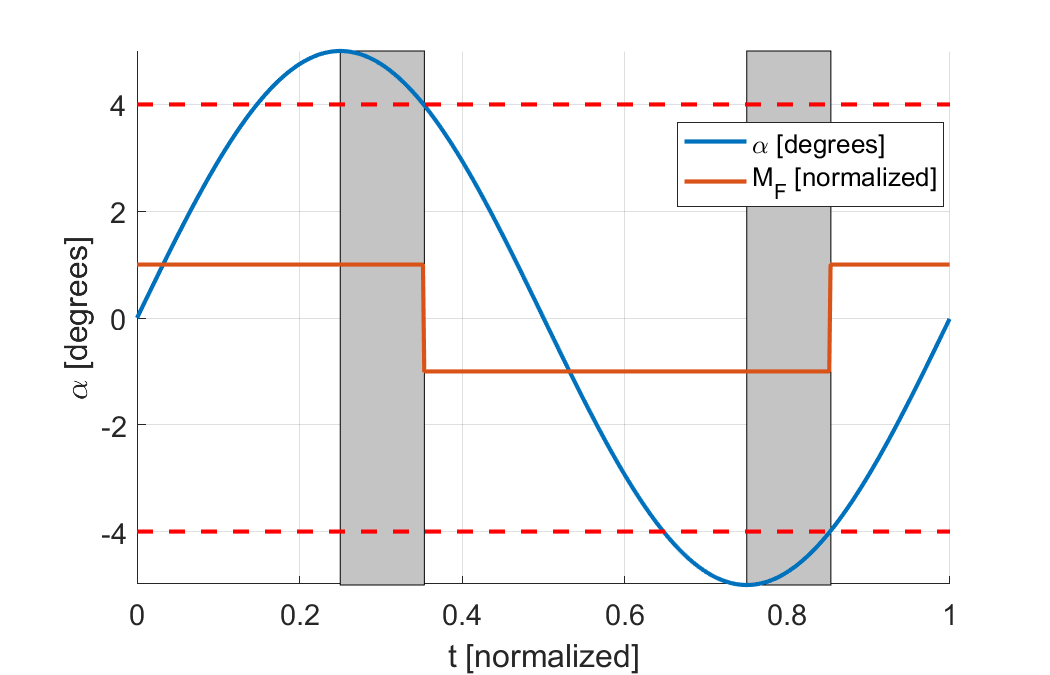}
\caption{Pendulum angle and driving force in grasshopper escapement model, as a function of time.}\label{fig:3}
\end{figure}

\begin{figure}[ht!]
\centering
a)\includegraphics[width=.85\linewidth]{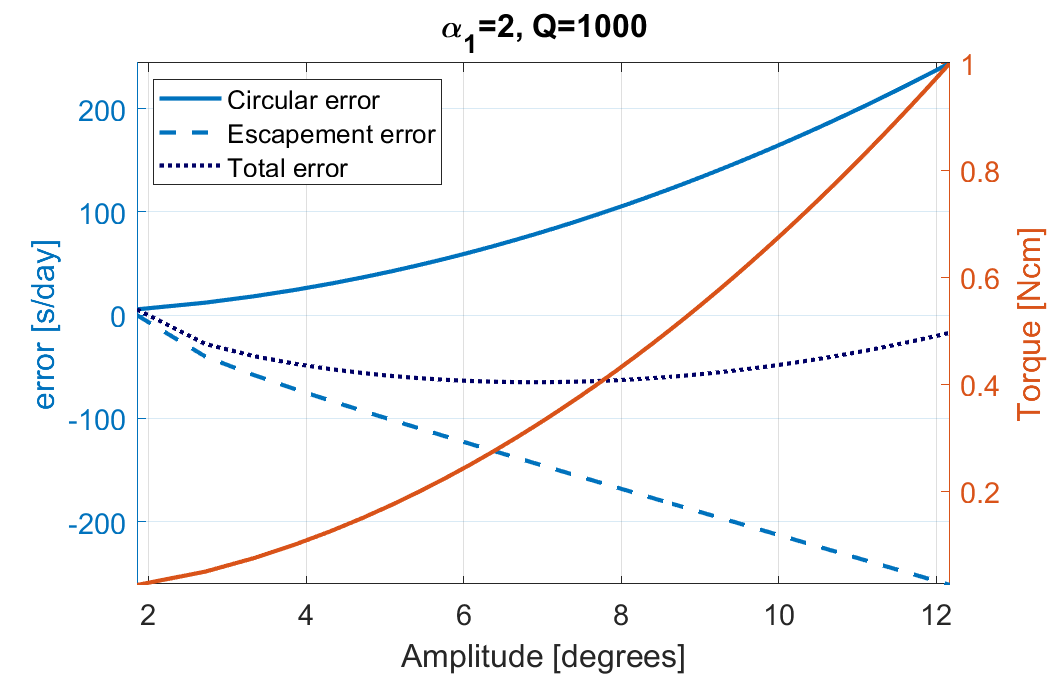}
b)\includegraphics[width=.85\linewidth]{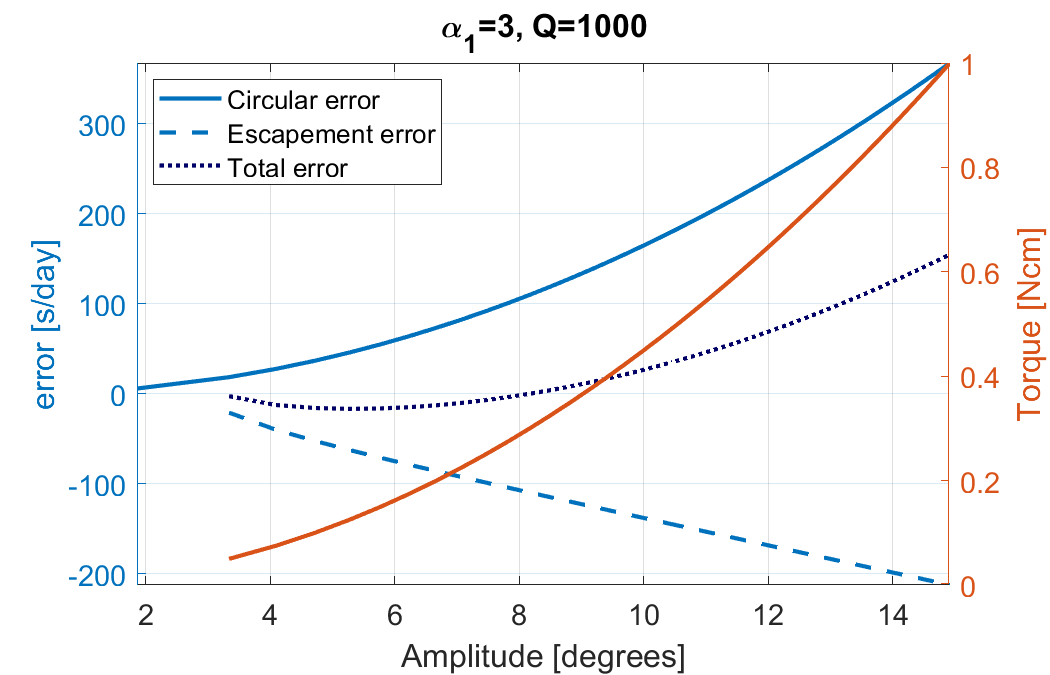}
c)\includegraphics[width=.85\linewidth]{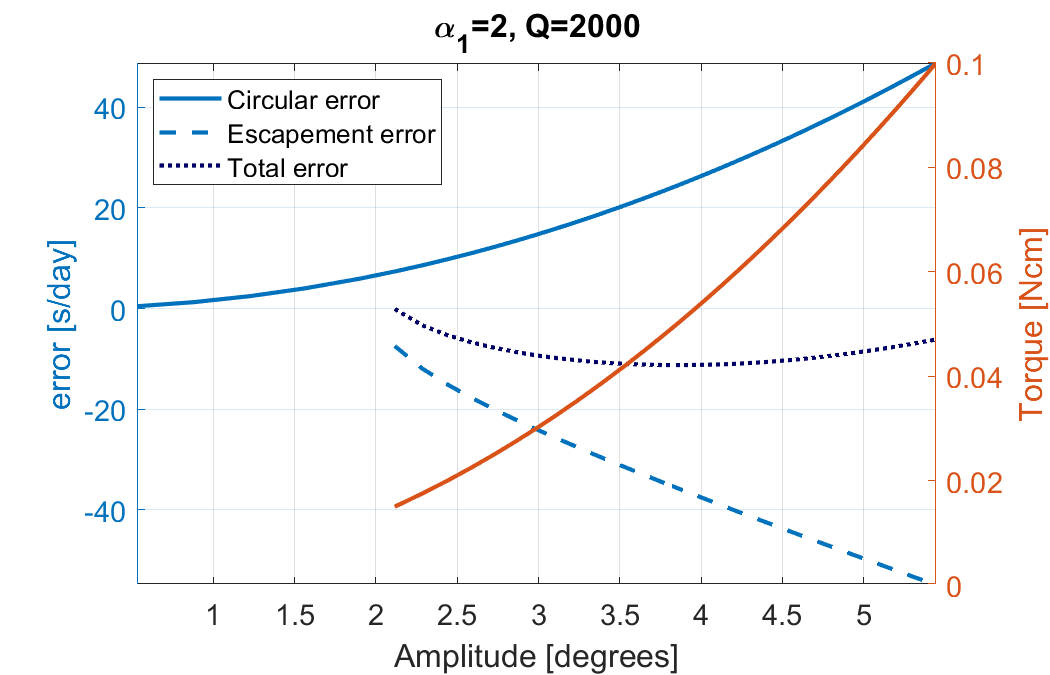}
\caption{Simulation results of grasshopper escapement with three different geometries.}\label{fig:4}
\end{figure}

For the purpose of analysis, the model constant-torque escapement used in \cite{Hoyng} will be adapted as a first approximation of the grasshopper escapement. The driving force is given by
\begin{equation}
M_F(\alpha)=M_{F0}\mbox{sgn}(\alpha+\alpha_1\mbox{sgn}(\dot{\alpha}))
\end{equation}
where $\mbox{sgn()}$ is the sign function. The above relation is shown on the Fig. \ref{fig:3}.
Initially, the angle $\alpha$ increases and the force $M_F$ is positive, pushing the pendulum in the direction of motion. After the maximum amplitude $A=5$ degrees is reached, the pendulum reverses direction while the force remains positive. In this region (marked by gray box) the mechanism exhibits recoil - the escapement is pushing against the pendulum. Finally, when $\alpha=\alpha_1=4$ degrees, the force switches sign. The cycle repeats in the second half of the period. Stable operation is possible only when $A>\alpha_1$; otherwise, the force never switches sign and the system stops. The system delivers energy to the pendulum when $\alpha_1>0$ (see Appendix B).
The simulation results obtained for various values of $\alpha_1$ and $Q$ are shown on the Fig. \ref{fig:4}.
Due to the fact that torque acts over whole pendulum motion instead of small angle limits, for any given $M_F$ the work done by the escapement and resulting amplitude in Fig. \ref{fig:4} a) is larger as compared to Fig. \ref{fig:2}. A very wide local minimum of error occurs at $\alpha \sim 7$ degrees. While the total error is significant over the whole amplitude range, its rate of change $\partial E_T/\partial A$ is very small at $A \in (6,8)$, making the system very tolerant to considerable changes of driving torque. By increasing the angle where recoil occurs to $\alpha_1=3$ degrees, one further increases the amplitude, as shown on the Fig. \ref{fig:4} b). Due to the fact that recoil occurs later, the $E_E$ is reduced and minimum of total error happens earlier, at $\alpha \sim 5$ degrees. However, a practical issue becomes apparent - the location of the minimum approaches the minimum amplitude $A=\alpha_1$ necessary for escapement operation. As in the case of the chronometer escapement, the escapement error scales with force and the location of the minimum depends on the ratio of $E_E$ to $E_C$. In order to shift the minimum towards smaller angles, one can significantly reduce the force and then increase the Q factor to reach the necessary minimal amplitude. Such a case is presented on the Fig. \ref{fig:4} c); the optimal working amplitude is $A \sim 4$ degrees and one can achieve about 2 seconds/day speed variation with $50\%$ torque variation (0.04 to 0.06 Ncm). By regulating the torque to fall within $1\%$ of nominal value, one can easily attain stability on the order of few seconds per year claimed in \cite{McEvoy}, provided that the compensation of environmental effects such as changes of temperature and air density is good enough. Note that in all above cases, the escapement exhibits considerable recoil - the amplitude is much larger than the angle $\alpha_1$, where the force switches direction. As mentioned in \cite{Harrison}, the recoil is not necessarily detrimental, especially in grasshopper escapement, where it produces very little additional friction.

The above results are consistent with general observation in \cite{Hoyng} that the escapement error has an opposite sign to the circular error and the relation between the two depends on the combination of driving force and friction. In general, while the circular error $E_C \sim A^2$, the escapement error is proportional to $A$ (See also Appendix B for an estimation of $E_E$). The above calculations indicate that the large pendulum amplitude is not always detrimental to the accuracy. In fact, when one takes into consideration the external disturbances from the environment, large $A$ may be beneficial. Due to the fact that pendulum energy $E \sim A^2$, the change of amplitude is $\partial A/\partial E \sim 1/\sqrt{E}$. For any given disturbance adding or subtracting some energy $\Delta E$, the pendulum with larger total energy is less affected. The same conclusion is provided by Harrison \cite{Harrison}. This problem is further explored in the next section. 

\begin{figure}[ht!]
\centering
\includegraphics[width=.9\linewidth]{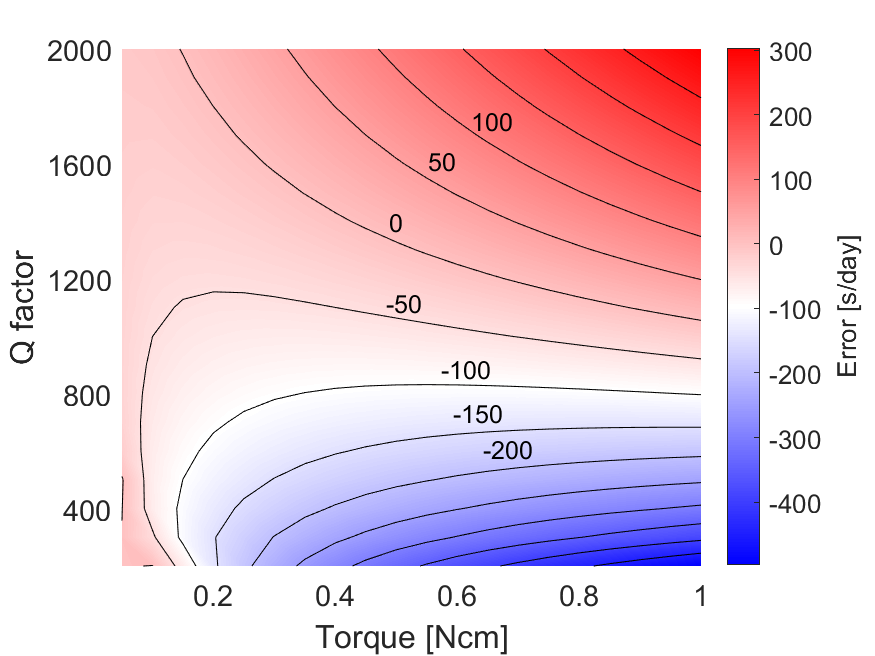}
\caption{The effect of variable torque and Q factor on the system from Fig. \ref{fig:4} a). Contours of constant error are marked with black lines.}\label{fig:5}
\end{figure}

The cancellation of the effects of $E_C$ and $E_E$ causing local insensibility to changes of amplitude does not occur when the said change is not caused by a change of torque. When the pendulum Q factor is decreased (due to the wear on the pendulum suspension or change of air density and thus resistance), the $E_C$ decreases due to the smaller amplitude and the $E_E$ also decreases (reaches larger negative value) due to the fact that the motion becomes less harmonic. Thus, both effects add up instead of compensating each other. The error calculated for a wide range of torque and Q factor is shown on the Fig. \ref{fig:5}. Overall, the error is negative ($|E_E|>|E_C|$) in the low Q region and positive for high Q. Note that the optimal conditions from the Fig. \ref{fig:4} a), i. e. $Q=1000$, $M_F\sim 0.4$ Ncm are located between $E_T=-100$ and $E_T=-50$ contour, inside a wide area where the error is insensitive to changes of torque, but changes rather quickly with $Q$. By changing the Q from 800 to 1200, one can expect a change of rate on the order of 50 seconds/day. This considerable error can be reduced by shifting the operating point towards lower Q factor; on the bottom left side of the Fig. \ref{fig:5}, the value of $\partial E_T/\partial Q$ is smaller (the contours are more vertical). However, at this operating point the mechanism is more sensitive to torque variation. Overall, the correct choice of the optimal driving torque depends not only on the escapement geometry and pendulum Q factor, but also on the expected variation of $M_F$ and $Q$. In this particular example, the smallest variation of error with both parameters occurs near $M_F=0.15$ Ncm, $Q=1500$. It should be noted that the balance between $E_C$ and $E_E$ and the resulting "landscape" of error can be further tuned by partially compensating for the circular error, for example by using a modification of Hyugens cycloid pendulum suspension \cite{Harrison}.

\section{Variable torque}
\begin{figure}[ht!]
\centering
\includegraphics[width=.9\linewidth]{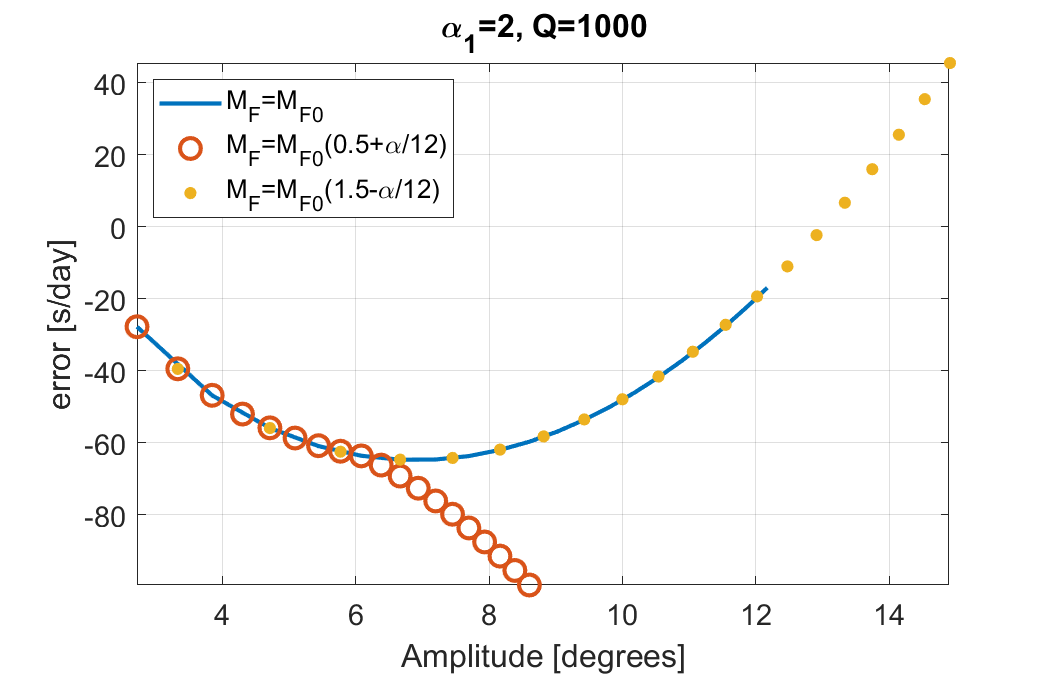}
\caption{The effect of variable torque on the system from Fig. \ref{fig:4} a).}\label{fig:5b}
\end{figure}

Practical realization of grasshopper escapement cannot guarantee exactly constant torque regardless of the pendulum angle. In order to study the effects of variable torque, the error for two different functions of $M_F(\alpha)$ is shown on the Fig. \ref{fig:5b}. 
In the first case, the torque is reduced to $50\%$ and increases linearly to $150\%$ at $\alpha=12$, which is the largest amplitude of the constant torque system. Several interesting effects occur; the increase of force near the swing ends reduces the amplitude and amplifies the escapement error. In the region where the maximum torque is smaller than $M_{F0}$ ($\alpha<6$), the error is almost exactly the same as in the constant force system. On the other hand, when the force is reduced at higher angles, the resulting amplitude becomes larger. In this case, the error is indistinguishable from the $M_F=\mbox{const}$ system. It should be noted that the dependence of amplitude on the function $M_F(\alpha)$ is nontrivial; for a given Q factor, the energy lost in one period depends on total energy $E \sim A^2$, which is then balanced with the work done by the escapement given by Eq. (\ref{eq:work}), that is also a function of amplitude.

The above results indicate that the system is mostly insensitive to variations of torque with angle as long as no significant increase of $M_F$ occurs near the swing end. Thus, the previous calculations performed with the constant torque model are quite general and applicable to real life systems. Furthermore, the calculations indicate that it is beneficial to design the escapement geometry in such a way that the torque acting on the pendulum reduces with its angle. 

The lack of sensibility to small changes of torque distribution is consistent with results of theoretical approaches such as in \cite{Atkinson}, where it is shown that the clock rate can be connected with integrals over the whole period under the assumption than to large changes of acceleration (and thus force) occur near the maximum swing angle. The large increase of error when significant force is applied near the amplitude is also consistent with original observations by Harrison \cite{Harrison}.

\section{Transient effects}
\begin{figure}[ht!]
\centering
a)\includegraphics[width=.84\linewidth]{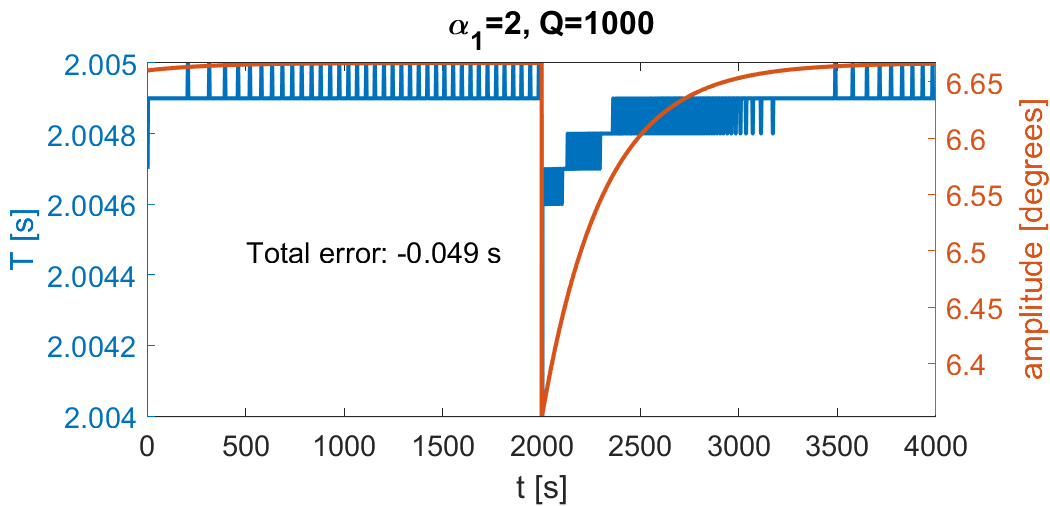}
b)\includegraphics[width=.84\linewidth]{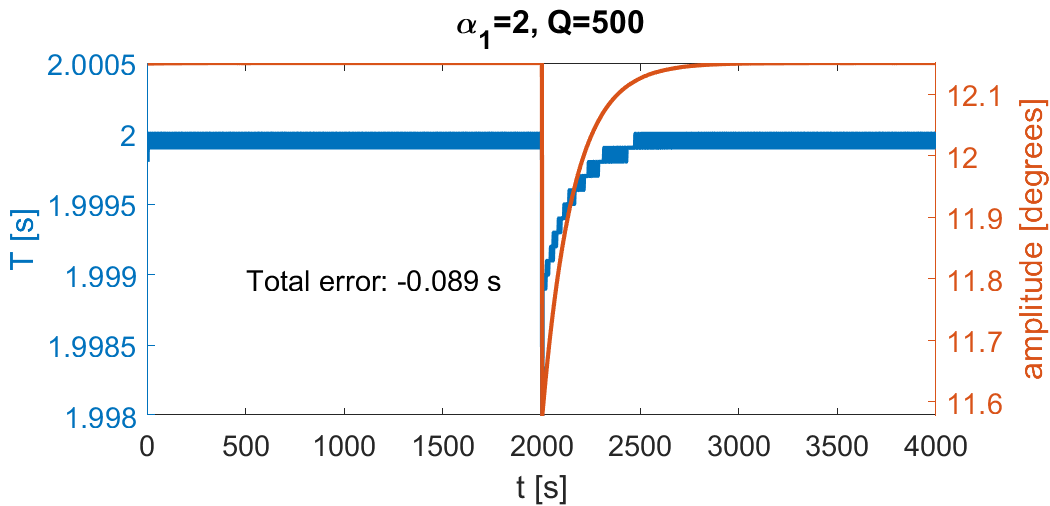}
c)\includegraphics[width=.84\linewidth]{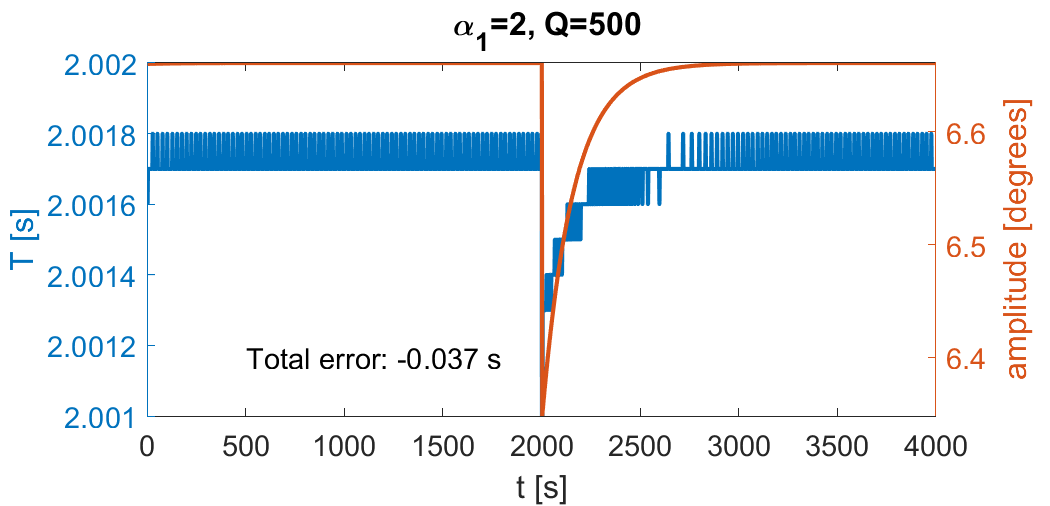}
\caption{Momentary change of amplitude and period of the pendulum subject to point disturbance at t=2000 s.}\label{fig:6}
\end{figure}

Lets consider the system presented on the Fig. \ref{fig:4} a), operating in the steady state regime. At some point, due to the ground vibration, sudden hit on the clock case etc. the momentary gravity acceleration is increased to $g'=2g$ for a period of $\Delta t=1$ ms. The phase of the pendulum at the moment when the disturbance occurs determines whether the period is increased or decreased. For example, one can choose the starting point of the disturbance to the moment when the pendulum passes the $\alpha=0$ angle. Then, the apparently increased gravity will slow down the pendulum on its way to $\alpha=A$. The computation results of such a setup are shown on the Fig. \ref{fig:6}.
The total error is calculated by integrating the momentary error ($T-T_{stationary}$) over time. On the Fig. \ref{fig:6} a) one can see the results for the system from the Fig. \ref{fig:4} a) operating at its optimal amplitude $A \sim 6.6$ degrees. After the initial shock, the system approaches the stationary condition exponentially, recovering almost completely over the time of 2000 s. The maximum change of the period is $\Delta T/T \sim 0.04 \%$ and the total error is on the order of 0.05 s. The results are in agreement with the well known general observation that the system averages out the very short-term errors and its dynamics is characterized by long term changes (in this case, noticeable effects can be observed over the time of several hours) \cite{Hoyng}. The Fig. \ref{fig:6} b) shows a modification of the system where the Q factor is reduced to 500. As mentioned before, in such a case the escapement error is greater and the local minimum of $E_T$ occurs at a higher angle of $A \sim 12.15$~degrees. Due to the larger amplitude and resulting circular error, the impact changes the rate more considerably, by $\Delta T/T \sim 0.1 \%$. However, the total error is only slightly larger due to the fact that a lower $Q$ pendulum approaches the steady state faster. In fact, for the same amplitude than in the first example, disturbance of the more damped pendulum yields lower total error (Fig.~\ref{fig:6}~c)). Again, it turns out that design goal of largest possible $Q$ factor is not always beneficial for clock accuracy.

\section{Chaotic motion}
All the above results are based on the assumption that the pendulum suspension point and escapement mechanism are designed in such a way that no significant mechanical vibrations occur in them during regular operation. The critical point in the motion of grasshopper mechanism where such vibrations may occur is the moment when $M_F$ changes sign; at this point, one of the pallets disengages from the escapement wheel, while the other impacts it. The stress wave propagates through the mechanism, exciting many vibrational modes in the structure \cite{Moon}.

\begin{figure}[ht!]
\centering
a)\includegraphics[width=.85\linewidth]{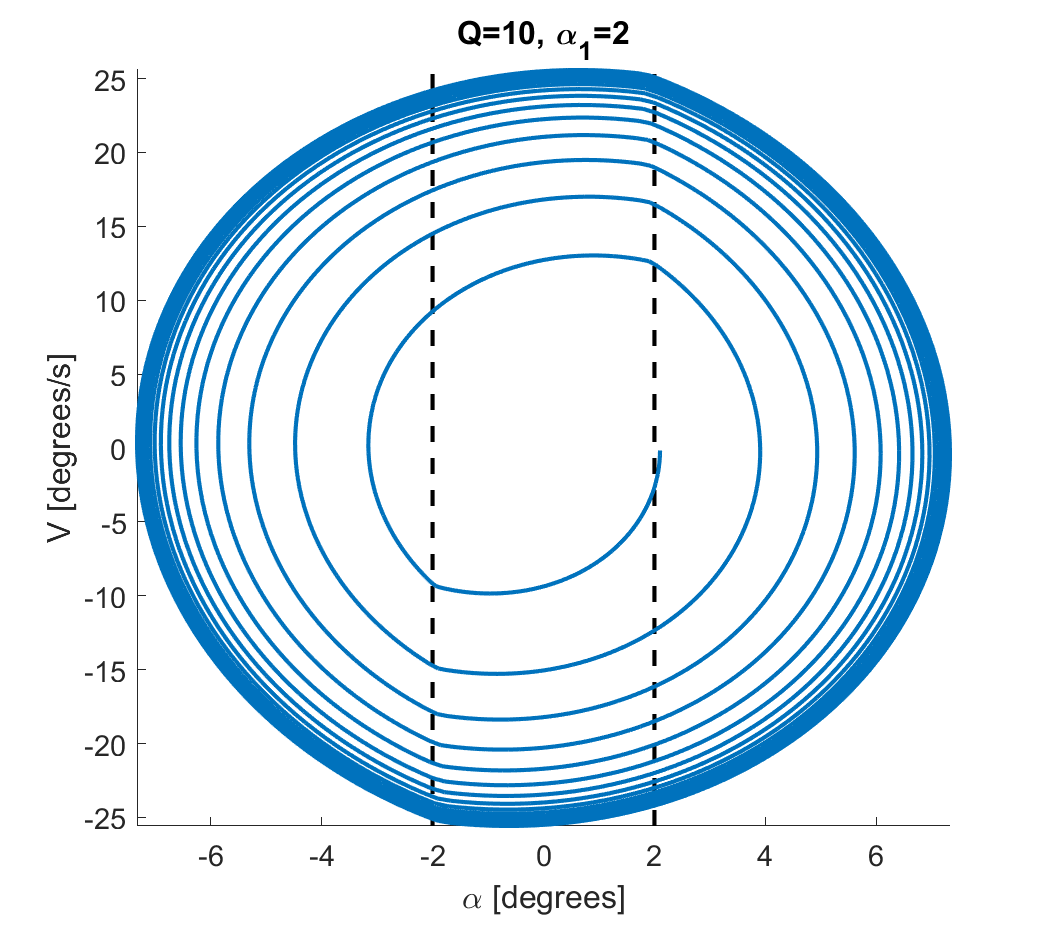}
b)\includegraphics[width=.85\linewidth]{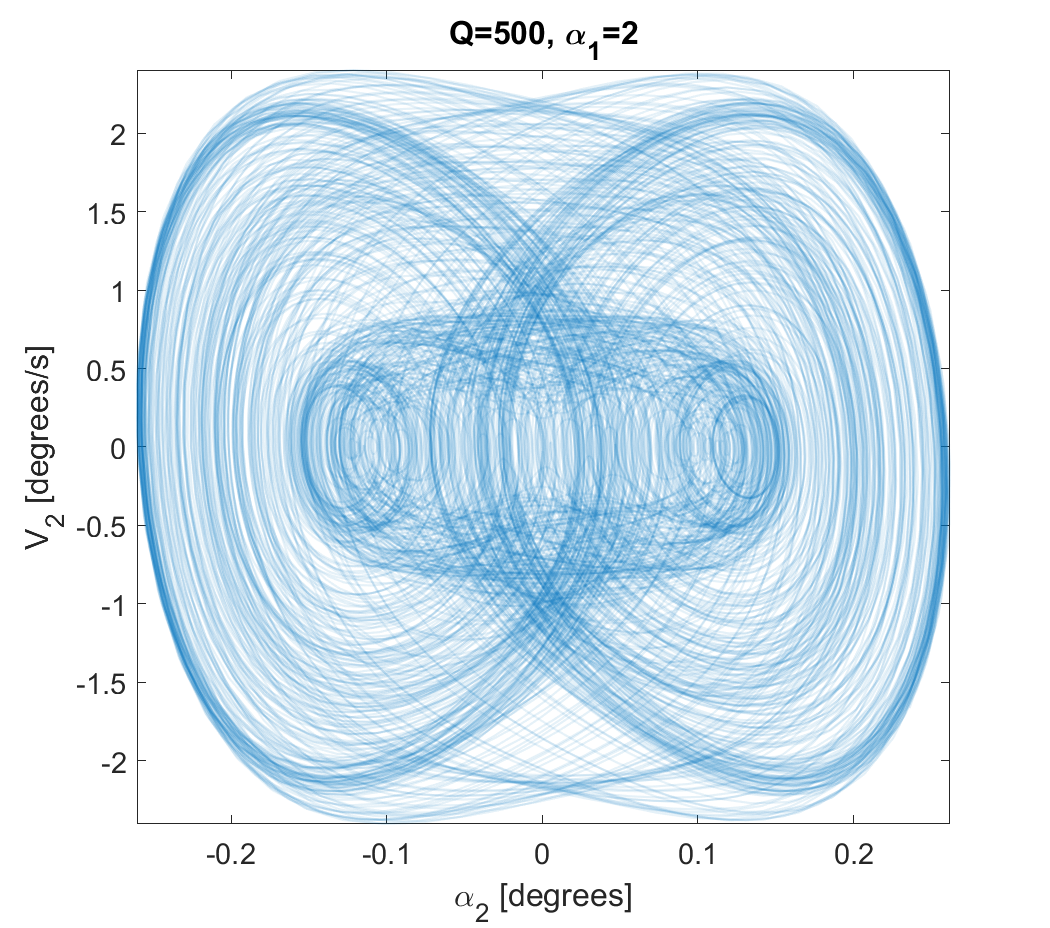}
\caption{a) Trajectory of the grasshopper escapement in the phase space. b) Trajectory of the second oscillator in the two-oscillator model.}\label{fig:7}
\end{figure}

\begin{figure}[ht!]
\centering
a)\includegraphics[width=.85\linewidth]{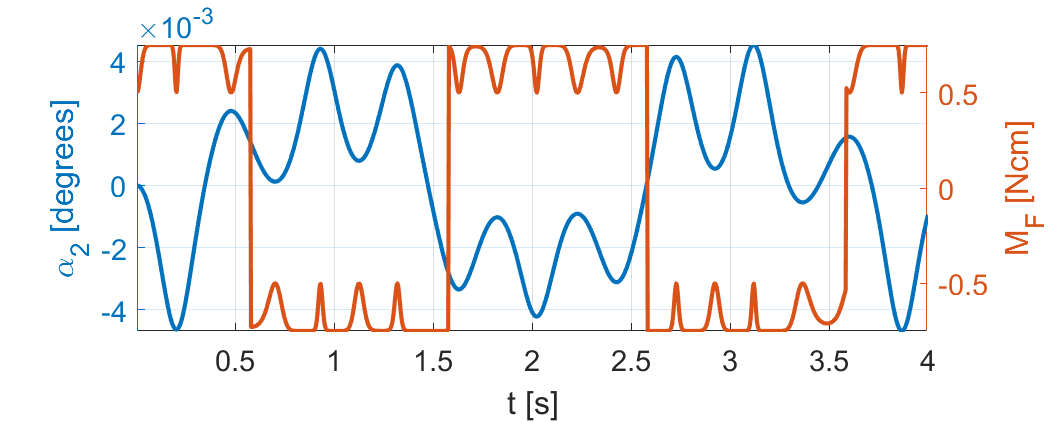}
b)\includegraphics[width=.85\linewidth]{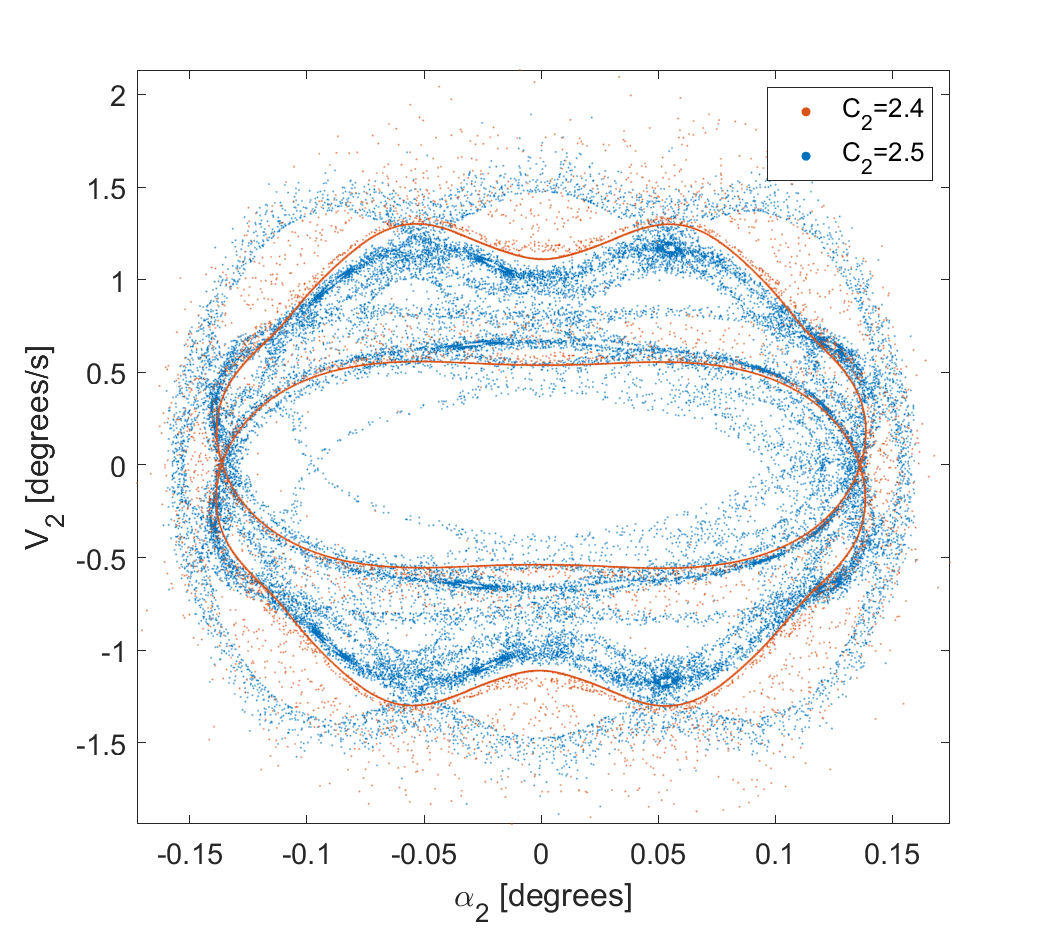}
\caption{a) Time evolution of $\alpha_2$ and $M_F$ b) Poincar\'{e} map of the system obtained for $\alpha=0$.}\label{fig:7b}
\end{figure}

First, lets consider a grasshopper escapement with a large driving torque $M_F=20$ Ncm and strongly damped pendulum. The trajectory in the phase space $(\alpha,V=\dot{\alpha})$ is shown in the Fig. \ref{fig:7} a).
One can see a spiral-like shape approaching the limit cycle representing a steady state amplitude of $A \sim 7$ degrees. Noticeable discontinuities occur at the above mentioned critical points $\alpha = \pm\alpha_1$. In the case of larger Q factor and smaller torque, the trajectory is more circular.
For the simulation of mechanism vibrations, one can adapt the model presented in \cite{Moon} by introducing a second harmonic oscillator, resulting in a system of equations of motion
\begin{eqnarray}
\ddot{\alpha} + \gamma_1\dot{\alpha} + \omega_{01}^2\alpha + C_1\alpha_2 = M_F(\alpha,\dot{\alpha_2})/mL^2,\nonumber\\
\ddot{\alpha_2} + \gamma_2\dot{\alpha_2} + \omega_{02}^2\alpha_2 + \kappa\alpha_2^3 + C_2\alpha = 0,
\end{eqnarray}
where $\gamma_1,\gamma_2$ are the damping constants, $\omega_{01},\omega_{02}$ are the natural frequencies of the oscillators, $\kappa$ is a nonlinear term \cite{Moon} and $C_1,C_2$ are coupling constants that connect the second oscillator with the pendulum. A simulation has been performed with the parameters $\gamma_1 = \pi/1000$ (pendulum $Q=1000$), $\gamma_2=\gamma_1/100$, $\omega_{02}=7\omega_{01}$, $\kappa=10^7$, $C_1=1$, $C_2=2.5$. Note that the large value of $\kappa$ is due to the fact that in computation, the angles are expressed in radians and thus for $\alpha_2 \sim 1$ degree, the $\alpha^3$ term is on the order of $10^{-6}$. The inclusion of this term means that the equation of motion for $\alpha_2$ is a Duffing oscillator \cite{Kao1987}, known for chaotic behaviour. The selected parameters serve as an exaggerated example and don't reflect any particular system; they were chosen under the general assumption that the mechanism vibrations are high frequency, highly nonlinear, weakly damped waves. The pendulum size and escapement geometry are the same as in the Fig. \ref{fig:3} a) and the nominal driving torque is $M_F=0.5$ Ncm; this value slowly increases with $\dot{\alpha_2}$, modelling reduced kinetic friction \cite{Moon}. To fully determine the state of the system, one needs two angular positions and two values of velocity/momentum. Thus, the full phase space of the system is 4-dimensional, e. g. $(\alpha,\dot{\alpha},\alpha_2,\dot{\alpha_2})$. By tracking only the second oscillator in its phase space $(\alpha_2,V_2=\dot{\alpha_2})$, one obtains results are shown on the Fig. \ref{fig:7} b). One can see a chaotic-looking trajectory resembling a strange attractor with two general groups of orbits characterized with positive and negative values of $\alpha_2$. This is confirmed by examining the evolution of $\alpha_2$ in time; on the Fig. \ref{fig:7b} a) one can see a square-like wave shifting between positive and negative values of $\alpha_2$, with additional, smaller, higher frequency oscillations. The shift between phases occurs roughly at the same moment as the shift of the torque. From the mechanical point of view, at this point one of the pallets disengages the escapement wheel while the other impacts it. Thus, any mechanical oscillations of the mechanism are most likely to be excited or affected at this moment. Note the small variation of the driving torque due to the above mentioned velocity-dependent friction. The trajectory in Fig. \ref{fig:7} b) is similar to the one obtained for Duffing oscillator driven by constant frequency source \cite{Kao1987}, which reflects the fact that the period of the pendulum is not significantly affected by oscillations of the mechanism. 
Finally, by taking a cross-section of the full phase space by selecting only the points where $\alpha=0$, one obtains a Poincar\'{e} map shown on the Fig. \ref{fig:7b} b). Two cases of $C_2=2.4$ and $C_2=2.5$ are shown. The increase of the coupling constant causes a sudden transition from a regular map to a self-similar, fractal-like one that exhibits symmetry breaking and period doubling \cite{Humieres}. In conclusion, the chaotic dynamics do not emerge unless the coupling constant is sufficiently large; an increase of $\gamma_2$ also has a stabilizing effect on the system. 

\begin{figure}[ht!]
\centering
a)\includegraphics[width=.85\linewidth]{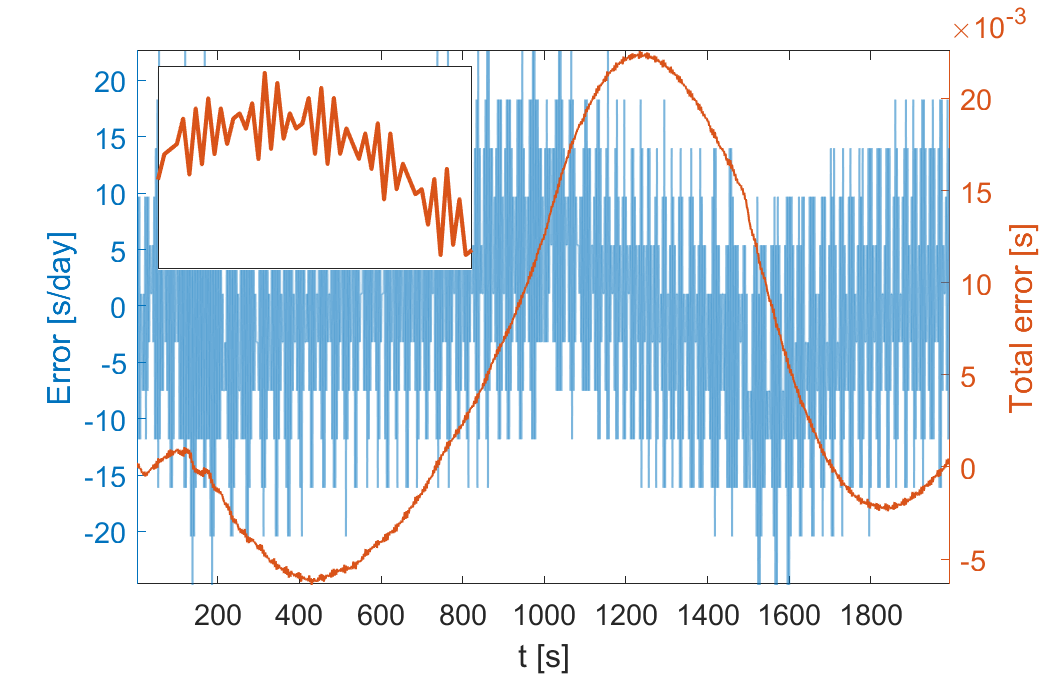}
b)\includegraphics[width=.85\linewidth]{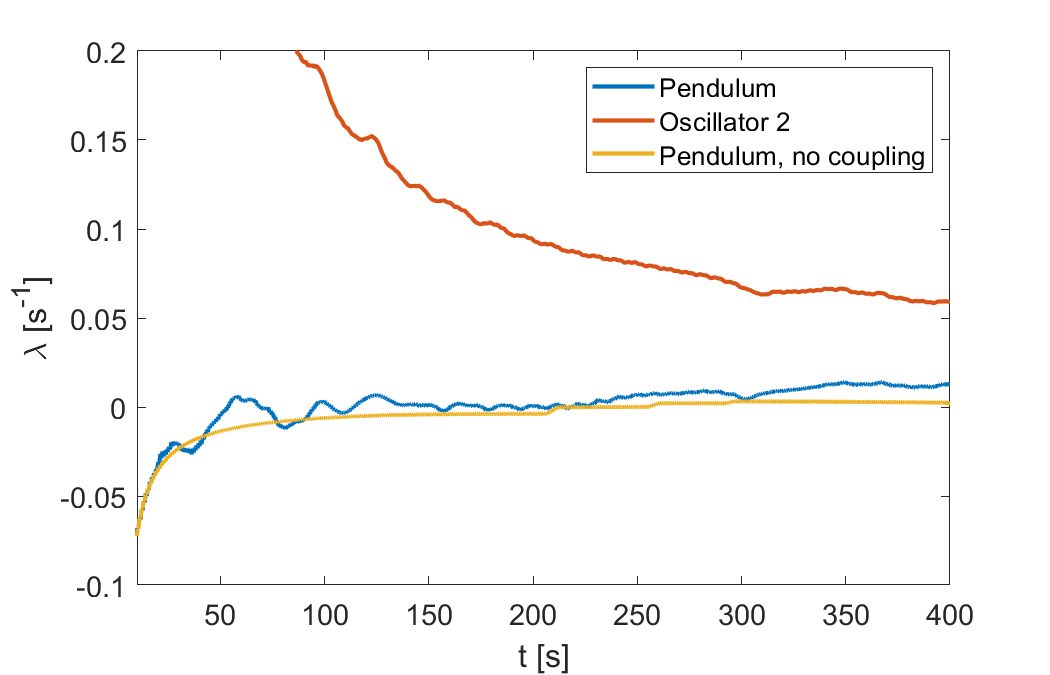}
\caption{a) Momentary error and total error due to the mechanism oscillations. b) Estimation of Lyapunov exponent of the system.}\label{fig:8}
\end{figure}

From the perspective of accurate timekeeping, the most important parameter is the overall effect of the vibrations on the pendulum speed, which is shown on the Fig. \ref{fig:8} a). 
One can notice a short-term, semi-random variation of the clock rate (blue line) on the order of 20 seconds/day. The apparent randomness is another indication that the system is chaotic. However, as mentioned in the discussion of transient effects, the final amplitude and period of a high-Q pendulum is a result of many small pushes from the mechanism delivered over considerable time, so that any short-time variation is averaged out. Therefore the total error (orange curve), which is a sum over momentary errors, is relatively insignificant. It is characterized by fractal-like structure containing both long-term and rapid oscillations (inset), similar to a random walk.

The key characteristic of a chaotic system is the fact that any two arbitrarily close starting points diverge exponentially in time; for example, for two pendulum simulations with trajectories $\alpha_1(t)$ and $\alpha_2(t)$, with slightly different starting condition $\alpha(0)$, one has
\begin{equation}
|\alpha_1(t)-\alpha_2(t)|\sim e^{\lambda t},
\end{equation}
where $\lambda$ is the so-called Lyapunov exponent. If $\lambda$ is positive, the system is chaotic \cite{Levien}. The value of $\lambda$ has been calculated by performing a simulation of two pendulums with an initial angle of $7$ and $7+10^{-6}$ degrees which, according to Fig. \ref{fig:7} c), is close to the steady state amplitude. The evolution of angle difference between the two pendulums is calculated and for every time $t$, an exponential function is fitted to the data, providing a set of values $\lambda(t)$. The results shown on the Fig. \ref{fig:8} b). The second oscillator (vibration of the mechanism) has a significant, positive exponent that in time converges to $\lambda \approx 0.05$. The value for the pendulum is much smaller and initially, it is negative. However, as the system approaches the maximum amplitude, the small variations induced by the mechanism become more pronounced, which results in a positive value $\lambda \approx 0.01$. Finally, when the coupling is turned off, the exponent converges to 0. This means that phase difference induced in the initial transient part of time evolution decreases exponentially over time, but two pendulums operating in the steady state regime will keep their phase difference constant over time. 

\section{Conclusions}
The chronometer and grasshopper escapement have been numerically studied. The changes of speed with variations of escapement geometry, driving torque and pendulum Q factor were investigated, showing that many theoretical results regarding long-term dynamics of the pendulum can be readily tested and confirmed with a direct numerical integration of equations of motion. The presented approach is both simple and flexible, allowing for study of steady-state motion as well as transient processes. It can be easily extended to model additional effects such as mechanical oscillations in the mechanism, leading to a chaotic motion. The results concerning the optimal working conditions of grasshopper escapement are in agreement with the Harrison's original notes and give some new insight into this peculiar system, often going against established guidelines applicable to other escapement mechanisms. It is shown that with careful selection of parameters, an exceptional accuracy can be achieved. The possibility of chaotic dynamics emerging in the system is investigated. 

\section*{Appendix A}
\begin{figure}[ht!]
\centering
\includegraphics[width=.8\linewidth]{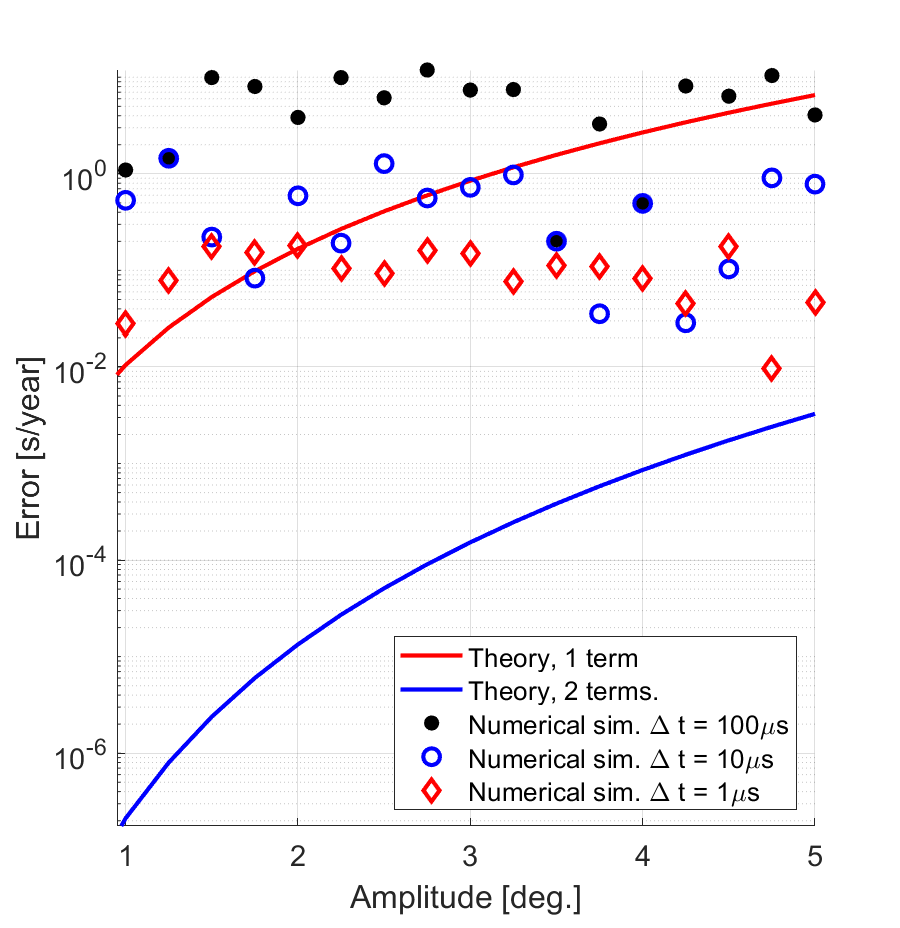}
\caption{Circular error obtained from simulations with various time steps.}\label{fig:a1}
\end{figure}

The pendulum motion is represented as a series of angle values $\alpha_n$ at times $t_n = n\Delta t$, where $\Delta t$ is a finite time step. The angle is calculated by integrating the equations
\begin{eqnarray}
V_{n+1}=V_n+\int\limits_t^{t+\Delta t}a(\tau) d\tau \approx V_n + a_t \Delta t,\nonumber\\
\alpha_{n+1}=\alpha_n+\int\limits_t^{t+\Delta t}V(\tau) d\tau \approx \alpha_n + V_t \Delta t,
\end{eqnarray}
where V is the angular velocity and the angular acceleration is given by
\begin{equation}
a = -\frac{g}{L}\sin(\alpha)-\frac{\gamma}{mL}V + \frac{M_F}{mL^2},
\end{equation}
where $\gamma$ is a damping constant. The drag force proportional to the velocity is a good approximation of the laminar drag \cite{Bolster}, provided that the velocity variation is small; in general, the drag coefficient reduces with Reynolds number, which is proportional to the velocity. Precise determination of drag forces can be achieved with fluid dynamics simulations \cite{Mongelli}.
 To choose the appropriate time step, a series of simulations has been performed (Fig. \ref{fig:a1}) and the difference between calculated circular error and the exact (within floating point arithmetic accuracy) value obtained with arithmetic-geometric mean \cite{Carwalhaes} has been investigated. Additionally, one- and two- term expansion of Eq. (\ref{eq:circ}) has been added. 
Interestingly, even one term expansion provides a good approximation of the circular error, with a difference of under 6 seconds/year (relative error $\sim 10^{-7}$) at $\alpha=5$ degrees. Numerical simulation spanning 2000 seconds, with a time step of 100 $\mu$s ($N=2\cdot 10^7$ steps in total) provides comparable accuracy when the period is averaged over the whole simulation time ($\sim 1000$ periods). By reducing the time step to 10 $\mu$s, for the same simulation time $N=2\cdot 10^8$ one obtains an error roughly 10 times smaller. However, the computation time, which is proportional to $N$, is increased by a factor of 10. Further reduction of time step to $\Delta t=1\mu$s provides only a modest increase of accuracy. Therefore, the $\Delta t=10\mu$s is selected.

In the simulations where a steady state has to be reached, considerable time on the order of hours is often needed for the amplitude to stabilize fully. In order to speed up such computations, a dynamic time step is used. The initial phase where the pendulum quickly increases its amplitude is simulated with $\Delta t =1$ ms. As the step number $n$ increases, the time step is reduced exponentially to its final value of $\Delta t=10\mu$s. By doing so, one can dramatically increase the total simulation time while preserving the small time step for the stationary regime where the changes of amplitude and period are very small and require additional accuracy. This method is used in all calculations regarding chronometer and grasshopper escapement, with $N=2\cdot 10^7$ steps and total time $t \approx 6000$ s. The steady-state period is an average of the last 100 values. 

\section*{Appendix B}
One can estimate the escapement error of the grasshopper escapement on the basis of energy conservation. Lets consider a single period that starts at $\alpha=-A$, $\dot{\alpha}>0$. The pendulum is constantly pushed in the direction of motion, passing the middle point $\alpha=0$ and climbing up to $\alpha=\alpha_1$, which results in the work done by the escapement
\begin{equation}
W_1 = M_F(A+\alpha_1),
\end{equation} 
where $M_F$ is assumed to be constant. In the last part of motion $\alpha_1<\alpha<A$ the torque counteracts the pendulum motion, doing work 
\begin{equation}
W_2 = -M_F(A-\alpha_1).
\end{equation}
These two phases repeat in the second part of the period. Neglecting the frictional losses, the kinetic energy increases by
\begin{equation}\label{eq:wtotal}
\Delta E_k = 2W_1+2W_2=4M_F\alpha_1.
\end{equation} 
Thus, a nonzero $\alpha_1$ is necessary to deliver energy to the pendulum.

As mentioned in the discussion of the chronometer escapement, a torque acting symmetrically around $\alpha=0$ has no impact on the period; therefore, for estimation of the escapement error, the crucial part of the pendulum motion is the recoil phase $\alpha>\alpha_1$. With this assumption, one can again consider the change of kinetic energy
\begin{equation}
\frac{\Delta E_k}{E_k} = \frac{2\Delta V}{V} = \frac{2M_F(A-\alpha_1)}{m\omega^2A^2} 
\end{equation} 
where $\omega$ is the angular frequency of the pendulum and $E_k=\frac{1}{2}m\omega^2A^2$ is the kinetic energy. With this, one obtains
\begin{equation}
\frac{T-T'}{T'} = \frac{V-V'}{V'} \approx \frac{\Delta V}{V} = \frac{M_F(A-\alpha_1)}{m\omega^2A^2}. 
\end{equation}
The perturbed period $T'$ is smaller than $T$, so that the error $E_E$ given by Eq. (\ref{eq:EE}) is negative. In a steady state, from Eq. (\ref{eq:Q}) and Eq. (\ref{eq:wtotal}) one has
\begin{equation}
M_F = \frac{\pi m \omega^2 A^2}{4Q\alpha_1}
\end{equation}
so that
\begin{equation}\label{eq:EE2}
E_E = \frac{-M_F(A-\alpha_1)}{m\omega^2A^2} = \frac{-\pi(A-\alpha_1)}{4Q\alpha_1}.
\end{equation}
The error has a linear dependence on $A$ and converges to $E_E(A=\alpha_1)=0$. It also decreases with increase of $\alpha_1$ and $Q$. The latter effect is an indirect result of smaller torque $M_F$ necessary to achieve given $A$ with larger $Q$. Despite simplistic nature of the derivation, the result reflects all key properties of escapement error visible on the Figs. \ref{fig:4}; the $E_E$ is a negative, linear function of $A$. By comparing Fig. \ref{fig:4} a) and Fig. \ref{fig:4} b) one can see that for any given $A$, the ratio of errors for $\alpha_1=2$ and $\alpha_1=3$ is close to $3/2$. The error for $Q=1000$ (Fig. \ref{fig:4} a)) is twice as large as for $Q=2000$ (Fig. \ref{fig:4} c)). For specific numerical example, lets consider the value of $E_E$ on the Fig. \ref{fig:4} a) for $A=5$ degrees, which is $E_E \approx -100$ seconds/day. From Eq. (\ref{eq:EE2}) one obtains $E_E \approx 1.178\cdot10^{-3} \approx 102$ seconds/day.

\end{document}